\documentclass[preprint,12pt]{elsarticle}

\usepackage[T1]{fontenc}
\usepackage[utf8]{inputenc}
\usepackage{amsmath,amssymb,bm,nccmath,braket,leftidx}
\usepackage{graphicx}
\usepackage{dcolumn,multirow,threeparttable,booktabs}
\usepackage{siunitx,enumitem,textcomp,pict2e}
\usepackage[dvipsnames]{xcolor}
\usepackage{float}
\usepackage{verbatim}
\usepackage[hidelinks]{hyperref}
\DeclareGraphicsExtensions{.pdf,.png,.jpg,.jpeg}
\biboptions{sort&compress}

\makeatletter
\def\ps@pprintTitle{%
  \let\@oddhead\@empty
  \let\@evenhead\@empty
  \let\@oddfoot\@empty
  \let\@evenfoot\@empty
}
\makeatother

\begin{document}

\begin{frontmatter}

\title{Chemical Substitution as a Route to Anomalous Surface Superconductivity in NiBi$_3$}

\author[inst1]{Anupam Barik}
\author[inst1]{Subhadip Chakrabarty}
\author[inst1]{Achintya Low\fnref{fn1}}
\author[inst2]{Sudip Pal}
\author[inst1]{Setti Thirupathaiah\corref{cor1}}
\cortext[cor1]{Corresponding author}
\ead{setti@bose.res.in}

\address[inst1]{Department of Condensed Matter and Materials Physics, S. N. Bose National Centre for Basic Sciences, Kolkata 700106, West Bengal, India}
\address[inst2]{UGC-DAE Consortium for Scientific Research, University Campus, Khandwa Road, Indore 452001, Madhya Pradesh, India}
\fntext[fn1]{Present address: Department of Physics,
Ben-Gurion University of the Negev,
Be'er-Sheva 84105, Israel.}

\begin{abstract}
\bf{Stabilizing two-dimensional superconductivity in a bulk superconducting material provides a unique route to exploring the interplay between superconducting condensates of different dimensionalities. Here, we present evidence for a distinct surface-associated superconducting state in Pb-substituted NiBi$_3$. NiBi$_{2.8}$Pb$_{0.2}$ exhibits a resistive onset at $T_{c,\mathrm{onset}}^{*}\approx8.1$~K, nearly twice the bulk superconducting transition temperature of approximately 4.1~K. The absence of a corresponding Meissner response or specific-heat anomaly at the higher-temperature transition supports its non-bulk origin. A cusp-like angular dependence of the critical field, well described by the two-dimensional Tinkham model, together with nonlinear current--voltage characteristics and signatures of a Berezinskii--Kosterlitz--Thouless transition, establishes the low-dimensional character of this superconducting state. Remarkably, the surface-associated superconductivity persists well beyond the bulk upper critical field, reaching a maximum ratio of $H_{c3}/H_{c2}\approx4.3$, far exceeding the conventional Saint-James--de Gennes value of 1.695. These results demonstrate that Pb substitution can stabilize an anomalously enhanced low-dimensional superconducting state in NiBi$_3$, providing a platform for investigating superconductivity at the boundary of a bulk superconductor.}
\end{abstract}

\end{frontmatter}

\section{Introduction}

Surface superconductivity, in which superconductivity remains confined to a near-surface region after bulk superconductivity has been suppressed, has been a subject of sustained interest since its early theoretical prediction within the Ginzburg--Landau (GL) framework~\cite{Ginzburg1964,ginzburg1964surface}. For a conventional superconductor with an ideal planar surface and a magnetic
field parallel to that surface, superconductivity can survive up to the surface critical field $H_{c3}$, which exceeds the bulk upper critical field $H_{c2}$. The standard GL model predicts $H_{c3}/H_{c2}\approx1.695$. Surface superconductivity is therefore particularly relevant for realizing high-field superconducting states and reduced-dimensionality phenomena in bulk materials~\cite{Song2025PRB}.

Recent developments have demonstrated that superconductivity can be induced or strongly modified within surface and interfacial regions. Electric-double-layer transistors, for example, can accumulate carriers electrostatically at a material surface and have enabled two-dimensional (2D) superconductivity in SrTiO$_3$, KTaO$_3$, PtBi$_2$, and ZrNCl~\cite{Ye2012,ueno2008electric,ueno2011discovery,ye2010liquid, Schimmel2024,Changdar2025}. Chemical doping and interfacial engineering provide complementary routes to such states. Hole doping in the topological insulator Sb$_2$Te$_3$ produces a surface superconducting transition near 9~K~\cite{zhao2015emergent}, whereas gold contacts on the Dirac semimetal Cd$_3$As$_2$ induce a superconducting surface state that has been associated with flat-band formation~\cite{shvetsov2019surface}.
Likewise, superconductivity in Pb--Bi$_2$Te$_3$--Pb heterostructures reveals multiple transitions arising from spatially distinct superconducting regions~\cite{qu2012strong}. These examples illustrate that surface and interface engineering can generate superconducting phases with properties that are markedly different from those of the corresponding bulk material.

On the other hand, $\mathrm{NiBi}_3$ is an orthorhombic, quasi-one-dimensional intermetallic superconductor with a bulk transition temperature of approximately 4.1~K. Previous studies have identified it as a weakly anisotropic, fully gapped, predominantly conventional $s$-wave superconductor~\cite{Fujimori2000,Gati2018,Zhao2018,Shang2023fully, Gangwar2026NiBi3}.  Although early reports proposed coexistence of magnetism and superconductivity, later investigations of high-quality crystals attributed the magnetic response primarily to extrinsic Ni impurities rather than intrinsic bulk ferromagnetism~\cite{Liu2020,pineiro2011possible,herrmannsdorfer2011structure}. Chemical substitution has previously been explored as a means of modifying its superconducting properties. For example, partial substitution of NiBi$_3$ with magnetic Co was reported to enhance the superconducting transition temperature to approx. 4.8~K for Co concentrations up to 4\%~\cite{jayaram1987superconductivity}. The superconducting properties of NiBi$_3$ are also strongly sensitive to external pressure. Further, the pressure can modify both the crystal structure and dimensionality, resulting in distinct quasi-one-dimensional NiBi$_3$, quasi-two-dimensional NiBi$_2$, and three-dimensional $\beta$-NiBi phases~\cite{clarke2019controlling}. Moreover, high-pressure measurements have shown that the superconducting transition temperature of NiBi$_3$ decreases monotonically with increasing pressure, reaching approximately 3~K at 2~GPa~\cite{gati2019effect}. The pronounced sensitivity of superconductivity to pressure therefore motivates the exploration of chemical substitution as an alternative means of modifying the electronic structure and superconducting properties of NiBi$_3$.

In this work, we investigate Pb-substituted $\mathrm{NiBi}_3$ single crystals, $\mathrm{NiBi}_{2.8}\mathrm{Pb}_{0.2}$, in which Pb is incorporated without a detectable structural phase transition. We find that Pb substitution produces an additional resistive transition at $T_{c,\mathrm{onset}}^{*}\approx 8.1$~K, well above the bulk transition at $T_c\approx 4.1$~K. The absence of corresponding magnetization and specific-heat anomalies at $T_{c,\mathrm{onset}}^{*}$ excludes a second bulk superconducting phase. Instead, the cusp-like angular dependence of $H_{c3}$, agreement with the 2D Tinkham model~\cite{Tinkham1996}, nonlinear $I$--$V$ characteristics, and Berezinskii--Kosterlitz--Thouless behavior support the existence of a two-dimensional surface superconducting state. This state remains stable well above $H_{c2}$, reaching an experimentally extracted $H_{c3}/H_{c2}\approx 4.3$, and establishes $\mathrm{NiBi}_{2.8}\mathrm{Pb}_{0.2}$ as a promising platform for investigating coexisting bulk and surface-confined superconductivity.

\section{Results and Discussion}

\subsection{Structural and Chemical Analysis}
\begin{figure}[htbp]\centering
\includegraphics[width=1\linewidth,keepaspectratio]{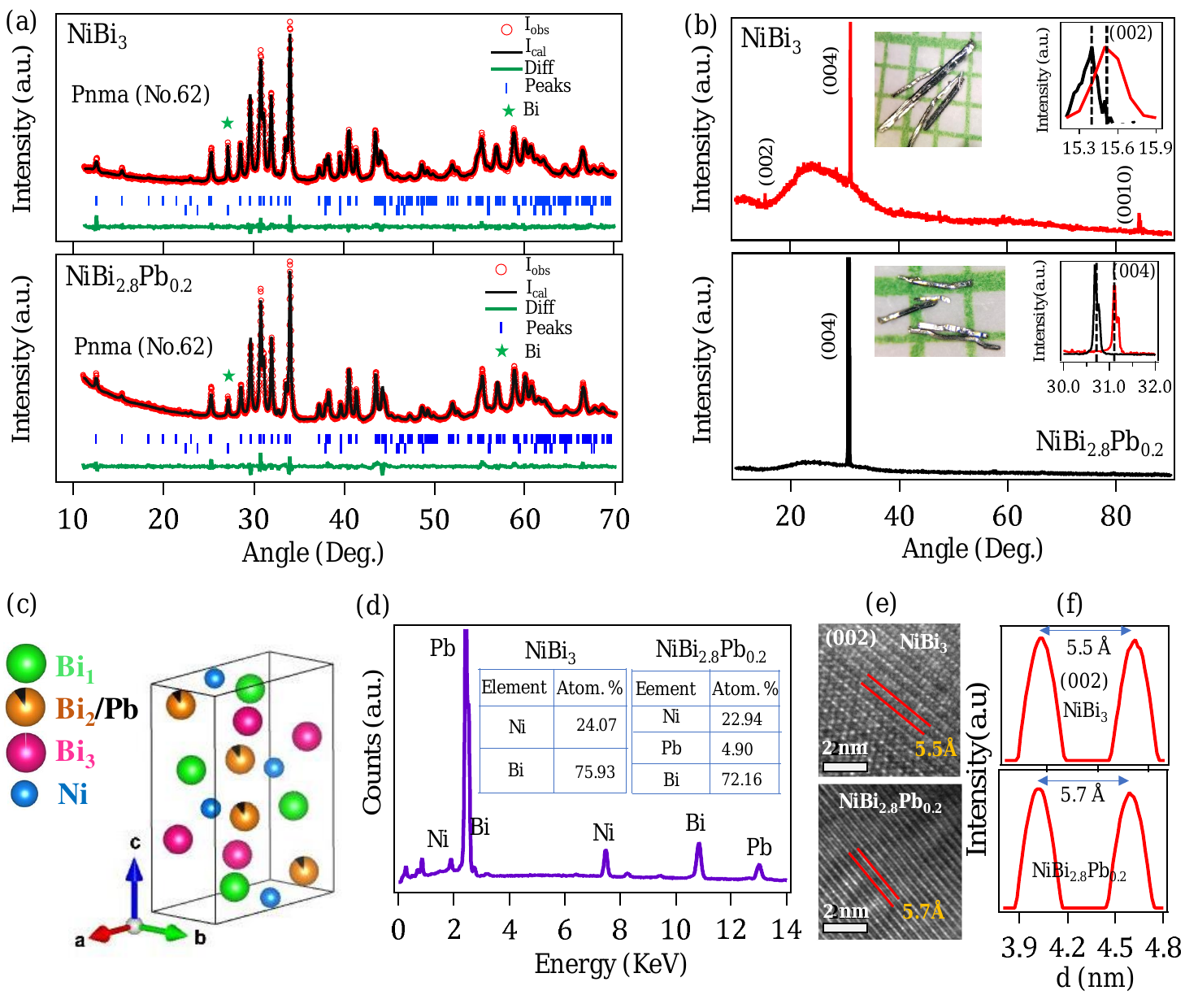}
\caption{(a) Powder X-ray diffraction (XRD) patterns and corresponding Rietveld refinements of crushed NiBi$_3$ and Pb-substituted ($\mathrm{NiBi}_{2.8}\mathrm{Pb}_{0.2}$) single crystals, shown in the top and bottom panels, respectively. (b) Single-crystal XRD patterns of NiBi$_3$ and $\mathrm{NiBi}_{2.8}\mathrm{Pb}_{0.2}$ crystals, shown in the top and bottom panels, respectively. The insets highlight the shifts of the ($002$) and ($004$) Bragg reflections between the parent and Pb-substituted crystals and include photographs of representative single crystals. (c) Schematic three-dimensional crystal structure of Pb-substituted NiBi$_3$, illustrating the partial replacement of Bi by Pb. (d) EDS spectra collected from Pb-substituted crystals, confirming Pb incorporation. The inset summarizes the atomic percentages of Ni, Bi, and Pb in the parent and Pb-substituted samples. (e) Enlarged HRTEM images of $\mathrm{NiBi}_3$ (top) and $\mathrm{NiBi}_{2.8}\mathrm{Pb}_{0.2}$ (bottom), showing the ($002$) lattice planes. (f) Corresponding lattice-fringe intensity profiles used to determine the ($d_{002}$) interplanar spacings.}
\label{fig.1}
\end{figure}

The chemical compositions of the as-grown single crystals were determined by energy-dispersive X-ray spectroscopy (EDS). As shown in Fig.~\ref{fig.1}(d), measurements on several Pb-substituted crystals yield an average composition of $\mathrm{Ni}_{0.92}\mathrm{Bi}_{2.88}\mathrm{Pb}_{0.20}$, whereas the parent crystals have a composition close to $\mathrm{Ni}_{0.96}\mathrm{Bi}_3$. These compositions are close to the nominal starting stoichiometries of $\mathrm{NiBi}_{2.8}\mathrm{Pb}_{0.2}$ and $\mathrm{NiBi}_3$, respectively, confirming the successful incorporation of Pb into the
$\mathrm{NiBi}_3$ crystals.

Fig.~\ref{fig.1}(a) presents the powder X-ray diffraction (XRD) patterns of crushed parent and Pb-substituted single crystals measured at room temperature, together with the corresponding Rietveld refinements. All major reflections of the $\mathrm{NiBi}_{2.8}\mathrm{Pb}_{0.2}$ can be indexed using the orthorhombic $\mathrm{NiBi}_3$ structure. The refinements confirm the centrosymmetric
orthorhombic structure with space group $Pnma$ (No.~62) for both NiBi$_3$ and $\mathrm{NiBi}_{2.8}\mathrm{Pb}_{0.2}$ samples~\cite{RODRIGUEZCARVAJAL1993,Fujimori2000,Kumar2011,silva2013superconductivity},
indicating that Pb substitution does not induce a structural phase transition in NiBi$_3$. For $\mathrm{NiBi}_{2.8}\mathrm{Pb}_{0.2}$, the refined lattice parameters are $a=8.890(2)$~\AA, $b=4.092(2)$~\AA, and $c=11.493(8)$~\AA. The corresponding lattice parameters of the parent compound are $a=8.879(6)$~\AA, $b=4.099(7)$~\AA, and $c=11.478(2)$~\AA~\cite{Liu2020,pineiro2011possible,herrmannsdorfer2011structure,Shang2023fully,Zhu2012surface}. Thus, the
orthorhombic lattice is retained upon Pb substitution, with only marginal changes in the lattice parameters. In particular, the $c$ axis increases by approximately $0.015$~\AA, consistent with a slight expansion along this crystallographic direction.

A weak additional reflection from elemental Bi is observed in the powder XRD pattern of both NiBi$_3$ and $\mathrm{NiBi}_{2.8}\mathrm{Pb}_{0.2}$. This feature is most likely associated with residual Bi flux adhering to the crystal surface, as frequently observed for $\mathrm{NiBi}_3$ single crystals grown by the Bi-flux method~\cite{Shang2023fully}. No additional diffraction peaks from Pb-containing impurity phases are detected within the resolution of the measurement. Together with the EDS results, the preserved crystal symmetry supports the incorporation of Pb into the $\mathrm{NiBi}_3$ host lattice without the formation of a detectable secondary
structural phase.

The effect of Pb substitution is also evident in the single-crystal XRD data shown in Fig.~\ref{fig.1}(b). The observed reflections are indexed according to the crystallographic $(00\it{l})$ orientation of the single crystals. The inset compares the $(002)$ and $(004)$ reflections of the parent and Pb-substituted crystals. The $(002)$ and $(004)$ peaks of the Pb-substituted sample shift towards lower $2\theta$, which according to Bragg's law indicates an increased interplanar spacing. This observation is consistent with the slight increase of the refined $c$-axis lattice parameter. The small structural change is reasonable given the similar atomic sizes of Bi and Pb.

A schematic of the orthorhombic $\mathrm{NiBi}_3$ crystal structure is shown in Fig.~\ref{fig.1}(c). The Ni atoms form zigzag chains along the crystallographic $b$ axis~\cite{Shang2023fully,Zhang2025nodal},  giving rise to the characteristic quasi-one-dimensional crystal morphology of $\mathrm{NiBi}_3$.  Further structural information is obtained from high-resolution transmission electron microscopy (HRTEM), as shown in Fig.~\ref{fig.1}(e). Well-resolved lattice fringes assigned to the $(002)$ planes are observed in both the parent
and Pb-substituted crystals. The lattice-fringe intensity profiles in Fig.~\ref{fig.1}(f) show a slightly larger apparent $(002)$ spacing in the Pb-substituted sample, consistent with the lower-$2\theta$ shift in XRD.

\subsection{Electrical Resistivity and Magnetotransport Properties}

\begin{figure}[htbp]\centering
\includegraphics[width=\linewidth,keepaspectratio]{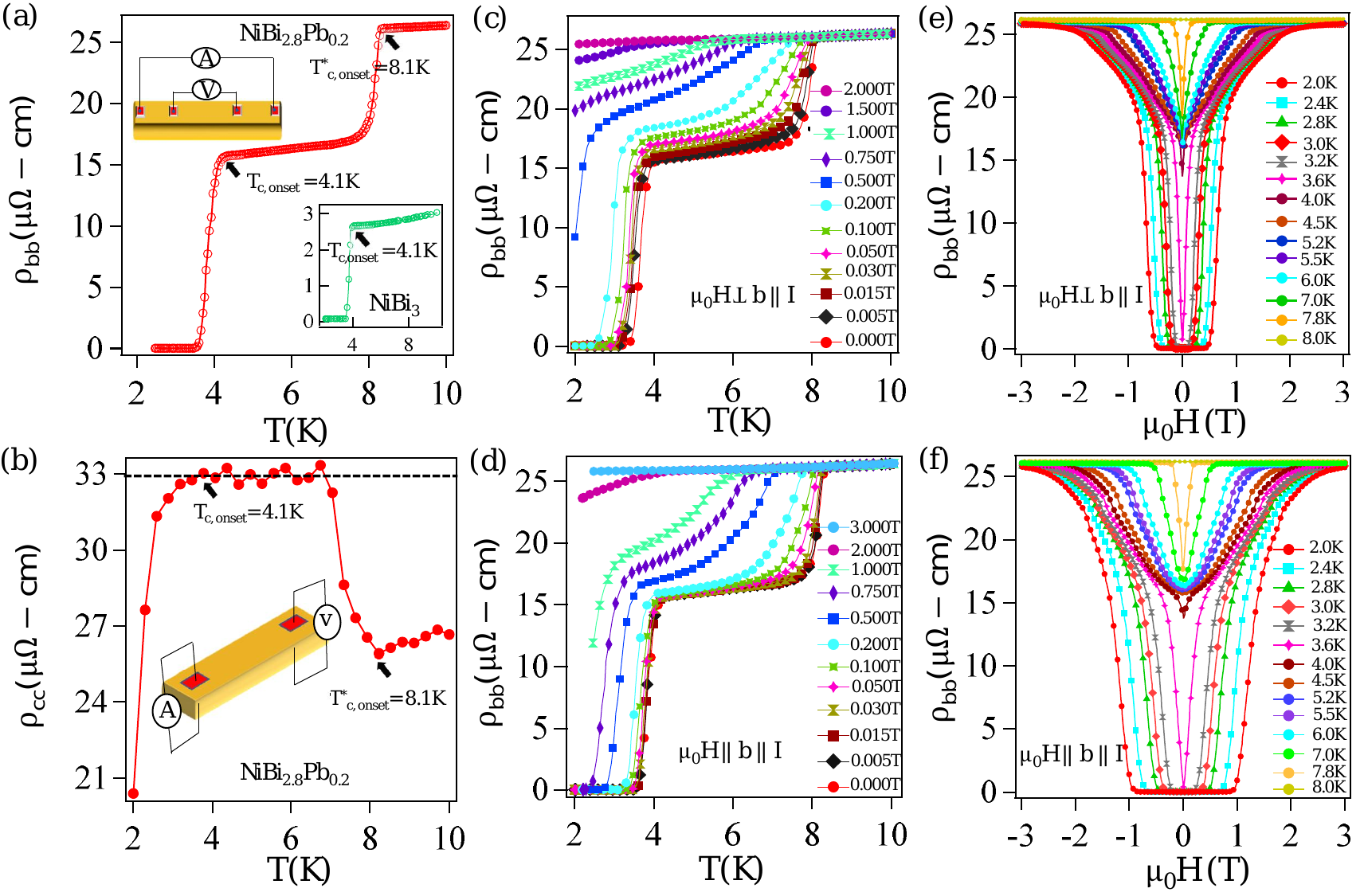}
\caption{(a) Temperature-dependent zero-field in-plane resistivity,$\rho_{bb}(T)$, measured with current applied along the crystallographic
$b$ axis for $\mathrm{NiBi}_{2.8}\mathrm{Pb}_{0.2}$. The lower-right inset shows $\rho_{bb}(T)$ for the parent $\mathrm{NiBi}_3$ compound, while the upper-left inset illustrates the measurement geometry. (b) Low-temperature out-of-plane resistivity, $\rho_{cc}(T)$, of
$\mathrm{NiBi}_{2.8}\mathrm{Pb}_{0.2}$, measured with current and voltage contacts along the crystallographic $c$ axis. (c) and (d) Temperature-dependent in-plane resistivity, $\rho_{bb}(T)$, measured under magnetic fields applied perpendicular ($H\perp b$) and parallel ($H\parallel b$) to the crystallographic $b$ axis, respectively. (e) and (f) Field-dependent in-plane resistivity, $\rho_{bb}(H)$, measured at various temperatures in the superconducting regime for $H\perp b$ and $H\parallel b$, respectively.}
\label{fig.2}
\end{figure}

Fig.~\ref{fig.2}(a) shows the temperature-dependent electrical resistivity of NiBi$_3$ and Pb-substituted $NiBi_{2.8}Pb_{0.2}$ single crystals. The parent NiBi$_3$ exhibits metallic behavior over the measured temperature range, followed by a sharp superconducting transition with an onset temperature of $T_{c,\mathrm{onset}}\approx4.1$~K, as shown in the inset of Fig.~\ref{fig.2}(a). This value is in good agreement with previously reported bulk superconducting transitions in NiBi$_3$~\cite{Fujimori2000,Gati2018,Zhao2018,Shang2023fully}. The sharp resistive transition, together with the previously reported Meissner response, establishes the bulk superconducting state of NiBi$_3$~\cite{Shang2023fully}.

In contrast, the Pb-substituted $NiBi_{2.8}Pb_{0.2}$ crystal exhibits a striking two-step resistive transition, indicating a substantial modification of the superconducting behavior upon Pb substitution. In addition to the bulk superconducting transition at $T_{c,\mathrm{onset}}\approx4.1$~K, a pronounced resistive drop emerges at a considerably higher temperature, with an onset at $T_{c,\mathrm{onset}}^{*}\approx8.1$~K. The appearance of this additional high-temperature resistive anomaly, which is absent in the parent compound, suggests the development of a distinct superconducting contribution induced by Pb substitution. The large separation between $T_{c,\mathrm{onset}}^{*}$ and the bulk $T_c$ further indicates that the two features are unlikely to represent a simple broadening of a single homogeneous superconducting transition. A possible origin of the higher-temperature transition is superconductivity confined predominantly to conducting planes, interfaces, or spatially restricted regions of the crystal. In such a scenario, superconducting correlations may first develop within highly conducting two-dimensional (2D) regions at $T_{c,\mathrm{onset}}^{*}$, producing a substantial reduction in the in-plane resistivity without immediately establishing three-dimensional (3D) phase coherence throughout the bulk. Upon further cooling, interlayer Josephson coupling and/or enhanced phase coherence between the superconducting regions may lead to a crossover from 2D to a fully coherent 3D superconducting state, giving rise to the lower-temperature bulk transition. Multistep resistive transitions associated with superconducting states of different effective dimensionality have also been reported in optimally doped Bi$_2$Sr$_2$CaCu$_2$O$_{8+\delta}$, where pressure-enhanced interlayer coupling drives a crossover between 2D and 3D superconducting behavior~\cite{Guo_2020}. Although the microscopic mechanism in $NiBi_{2.8}Pb_{0.2}$ may be different, this comparison provides a useful framework for examining whether the two resistive transitions originate from superconducting states with different dimensional characteristics.

To further probe this possibility, Fig.~\ref{fig.2}(b) presents the temperature dependence of the out-of-plane resistivity, $\rho_{cc}(T)$. The schematic configurations of the current and voltage contacts used for the in-plane and out-of-plane transport measurements are shown in Figs.~\ref{fig.2}(a) and \ref{fig.2}(b), respectively. A pronounced anisotropy is evident when the two transport geometries are compared. Whereas the in-plane resistivity displays a clear and substantial decrease beginning near $T_{c,\mathrm{onset}}^{*}\approx8.1$~K, $\rho_{cc}(T)$ shows only a weak resistive anomaly in the same temperature range. Importantly, the out-of-plane resistivity remains finite over a broad temperature interval and does not reach zero down to our lowest measured temperature of 1.8~K. The markedly different responses of the in-plane and out-of-plane resistivities therefore suggest that the superconducting correlations emerging near 8.1~K do not immediately establish coherent superconducting transport along the crystallographic $c$ axis. Such behavior is qualitatively consistent with a superconducting state that initially develops with a predominantly two-dimensional character~\cite{Li2007PRL, Guo_2020}. In this picture, superconducting pairing and/or phase coherence first becomes established within the $ab$ planes near $T_{c,\mathrm{onset}}^{*}$, strongly suppressing the in-plane resistance, while comparatively weak interlayer coupling prevents simultaneous coherent transport along the $c$ direction. With decreasing temperature, the interlayer coherence progressively strengthens, eventually producing a bulk three-dimensional superconducting state at lower temperature. The delayed development of zero resistance in $\rho_{cc}(T)$ relative to the high-temperature anomaly in the in-plane resistivity therefore provides transport evidence for strongly anisotropic superconducting coherence in Pb-substituted $NiBi_{2.8}Pb_{0.2}$.

This pronounced difference between the in-plane and out-of-plane transport provides important evidence for the low-dimensional character of the high-temperature superconducting state. If the 8.1~K transition originated from a homogeneous three-dimensional bulk superconducting phase, a pronounced similar resistive anomaly would be expected in both $\rho_{bb}(T)$ and $\rho_{cc}(T)$. Instead, the high-temperature transition is clearly manifested in the in-plane resistivity while producing only a weak feature in the out-of-plane resistivity. A superconducting layer located on the crystal surface and extending primarily parallel to the $ab$ plane can efficiently short-circuit the in-plane current, producing a pronounced reduction in $\rho_{bb}$. In contrast, the out-of-plane current must traverse the normal-state bulk and is therefore much less sensitive to a thin surface superconducting layer. Similar contrasting in-plane and out-of-plane transport signatures have previously been associated with two-dimensional superconductivity in layered superconductors~\cite{Li2007PRL}.  The microscopic role of Pb in generating this high-temperature superconducting state remains an open question. To assess whether Pb-containing systems can intrinsically support superconductivity at comparable temperatures, we examined previous reports on Pb--Bi alloys and Pb--Bi thin films. The Pb$_7$Bi$_3$ phase exhibits a bulk superconducting onset temperature of approximately $9$~K~\cite{karn2024superconductivity}, while the Pb$_{0.64}$Bi$_{0.36}$ alloy has been reported to show superconductivity at $T_c\approx8.6$~K~\cite{gutowska2024anisotropic}. Further, the superconductivity in two-monolayer, highly crystalline Pb$_{1-x}$Bi$_x$(111) films can be tuned continuously with Bi concentration, with $T_c=6.6$~K for $x=0.25$ and $T_c=4.7$~K for $x=0.5$~\cite{xie2023tuning}. In these Pb$_{1-x}$Bi$_x$ films, Bi substitution substantially modifies the electronic structure and electron--phonon coupling (EPC). At low Bi concentrations, the enhanced contribution of Pb-derived electronic states near the Fermi level strengthens the EPC and increases $T_c$, whereas further Bi substitution reduces the EPC strength and suppresses superconductivity~\cite{xie2023tuning}. These results suggest that Pb--Bi chemistry can support superconductivity in the temperature range relevant to the high-temperature transition observed here and suggest that modifications of the electronic structure and EPC may play a role in Pb-substituted NiBi$_3$. However, establishing such a microscopic mechanism in the present system requires further electronic-structure and lattice-dynamic investigations.

To further establish the superconducting nature of the high-temperature transition, we performed temperature-dependent resistivity measurements under applied magnetic fields. Measurements were carried out with the magnetic field oriented both perpendicular ($H\perp b$) and parallel ($H\parallel b$) to the crystallographic $b$ axis, while the electrical current was always applied along the $b$-axis. As shown in Fig.~\ref{fig.2}(c), the bulk superconducting transition near 4.1 K systematically shifts toward lower temperatures for $H\perp b$ ~\cite{Liu2025PRB,Shang2023fully}. At the same time, the high-temperature resistive transition is progressively suppressed and disappears at approximately 2~T. A similar behavior is observed for $H\parallel b$  as shown in Fig.~\ref{fig.2}(d), although the high-temperature transition persists to a higher field and disappears at approx. 3~T. The systematic suppression of the high-temperature resistive feature by magnetic field therefore provides further evidence that the 8.1~K transition is superconducting in origin~\cite{Guo_2020}. The applied magnetic field also produces a progressive broadening of the superconducting transitions. Such field-induced broadening is commonly associated with thermally activated flux flow (TAFF) in type-II superconductors and has been observed in several high-T$_c$ superconductors such as iron-pnictides, iron-chalcogenides, and cuprates~\cite{Gurevich_2011,Xing_2018,Ayas_2011,Shahbazi_2015}.

The field-dependent resistivity further reveals the distinct evolution of the superconducting states, as shown in Figs.~\ref{fig.2}(e) and \ref{fig.2}(f). Figure~\ref{fig.2}(e) presents $\rho_{bb}(\mu_0H)$ at temperatures ranging from 2 to 8.4~K for $H\perp b$. Below 4~K, a well-defined zero-resistance plateau is observed around zero field. With increasing magnetic field, the resistivity rises sharply, marking the suppression of the bulk superconducting state at the upper critical field ($H_{c2}$)~\cite{Zeinali_2016,Purwar_2024}. A broader resistive tail persists beyond this field, indicating an additional superconducting contribution that survives to higher fields and can be associated with surface superconductivity~\cite{Zeinali_2016}. Upon increasing temperature, the zero-resistance region progressively shrinks and vanishes near the bulk $T_c\approx4.1$~K. Above this temperature, the resistivity develops a pronounced minimum around zero field, which gradually weakens with increasing temperature and is suppressed by fields of order 3~T. The persistence of this field-sensitive feature above the bulk $T_c$ demonstrates that the superconducting response extends well beyond the bulk transition. The corresponding surface critical fields extracted from these resistive features are discussed in the later sections.

\begin{figure}[htbp]\centering
\includegraphics[width=0.5\linewidth]{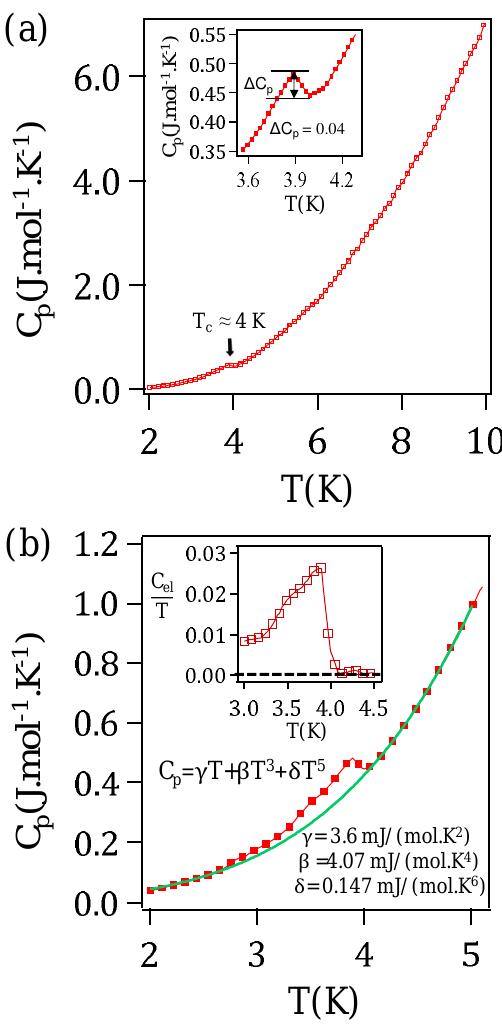}
\caption{(a) Zero-field heat capacity [$C_P(T)$] of $\mathrm{NiBi}_{2.8}\mathrm{Pb}_{0.2}$ single crystals as a function of
temperature. The inset enlarges the superconducting anomaly and indicates the heat-capacity jump [$\Delta C_P$] at the bulk transition temperature. (b) Low-temperature heat capacity of $\mathrm{NiBi}_{2.8}\mathrm{Pb}_{0.2}$. The solid green curve is a fit using
$C_P=\gamma T+\beta T^3+\delta T^5$, where $\gamma T$ is the electronic contribution and the higher-order terms describe the phonon contribution. The inset shows the electronic heat capacity [$C_{\mathrm{el}}/T$] as a function of temperature.}
\label{fig.3}
\end{figure}

For $H\parallel b$, the overall field-dependent behavior is similar to that observed for $H\perp b$, as shown in Fig.~\ref{fig.2}(d). Below approximately 4~K, the zero-resistance region persists over a wider field range than that observed for $H\perp b$, indicating a comparatively larger bulk upper critical field for $H\parallel b$. Above the bulk $T_c$, the field-sensitive resistance minimum remains visible over a finite field range and gradually weakens with increasing temperature, indicating the persistence of the high-temperature superconducting state. The reproducibility of the two-step transition in repeated electrical-transport and magnetic measurements on multiple crystals from both the same and different growth batches confirms that the high-temperature feature is intrinsic to the Pb-substituted NiBi$_3$ samples rather than an isolated sample-dependent effect. Note that we successfully reproduced the two-step resistive transitions through repeated electrical transport measurements on multiple crystals from both the same batch and different batches, confirming that the high-temperature feature is intrinsic to the Pb-doped samples rather than an isolated sample-dependent effect [see Supplemental Material for more details].

\subsection{Heat Capacity}

After confirming the emergence of high-temperature superconductivity with an onset transition temperature of $T_{c,\mathrm{onset}}^{*}=8.1$~K induced by Pb doping in NiBi$_3$, which we attribute to surface superconductivity, we performed specific-heat measurements to investigate the nature of the superconducting states. The zero-field specific-heat data exhibit a clear anomaly with a significant jump of $\Delta C_p = 0.038~\mathrm{J\,mol^{-1}\,K^{-1}}$ around $T\approx4\,\mathrm{K}$, as shown in Fig.~\ref{fig.3}. This anomaly is consistent with the bulk superconducting transition observed in both magnetization and resistivity measurements~\cite{Purwar_2024,Ma2022Ba5In4Bi5}. In contrast, no discernible anomaly is observed near the higher-temperature transition at $T_{c,\mathrm{onset}}^{*}=8.1$~K, which is clearly identified in the $\rho_{bb}(T)$ data. The absence of a corresponding specific-heat anomaly at this temperature strongly suggests that the superconductivity emerging at $8.1\,\mathrm{K}$ does not represent a bulk superconducting transition.

The experimental specific-heat data of NiBi$_{2.8}$Pb$_{0.2}$ were fitted in the temperature range of 4.0--5.0~K using the expression~\cite{Song2025PRB,Fujimori2000,Yan2019PRB}
\begin{equation}
C_p = \gamma T + \beta T^3 + \delta T^5,
\end{equation}
where $\gamma T$ represents the electronic contribution, while the $\beta T^3$ and $\delta T^5$ terms describe the phonon contributions. From the fit, shown by the solid green line in Fig.~\ref{fig.3}(b), we obtain $\gamma=3.6~\mathrm{mJ\,mol^{-1}\,K^{-2}}$, $\beta=4.07~\mathrm{mJ\,mol^{-1}\,K^{-4}}$, and $\delta=0.147~\mathrm{mJ\,mol^{-1}\,K^{-6}}$. The inset of Fig.~\ref{fig.3}(b) shows the electronic specific heat divided by temperature [$C_{\mathrm{el}}/T$] plotted as a function of temperature. A pronounced peak is observed in $C_{\mathrm{el}}/T$, providing clear evidence for the bulk superconducting transition. The electronic contribution is obtained by subtracting the phonon contribution from the total specific heat according to
\begin{equation}
C_{\mathrm{el}} = C_p-(\beta T^3+\delta T^5).
\end{equation}

The superconducting transition temperature, $T_c$, is estimated to be 3.95~K using the equal-area entropy construction~\cite{Yan2019PRB,Shen_2020}. This value is in good agreement with the transition temperatures obtained from the resistivity and magnetic susceptibility measurements. The normalized specific-heat jump, $\Delta C_p/\gamma T_c$, is estimated to be approx. 2.67, which is higher than the Bardeen--Cooper--Schrieffer (BCS) weak-coupling value of 1.43~\cite{Yan2019PRB, Shen_2020}. This enhanced specific-heat jump provides further evidence for the bulk superconducting nature of the transition near 4~K and may indicate deviations from the conventional weak-coupling BCS limit.

The Debye temperature, $\Theta_D$, can be estimated from the phonon coefficient $\beta$ using
\begin{equation}
\Theta_D = \left(\frac{12\pi^4 nR}{5\beta}\right)^{1/3}=124~\mathrm{K},
\end{equation}
where $n$ is the number of atoms per formula unit and $R$ is the universal gas constant. The obtained electronic specific-heat coefficient, $\gamma$, and Debye temperature, $\Theta_D$, for Pb-substituted NiBi$_3$ are lower than the corresponding values of $12.7~\mathrm{mJ\,mol^{-1}\,K^{-2}}$ and 144~K, respectively, reported previously for parent NiBi$_3$~\cite{Fujimori2000}. The absence of a detectable specific-heat anomaly at 8.1~K, together with the pronounced bulk anomaly near 4~K, provides compelling evidence for the coexistence of high-temperature surface superconductivity with bulk superconductivity below 4~K. A similar coexistence of surface- and bulk-related superconducting behavior has been reported in optimally doped Bi$_2$Sr$_2$CaCu$_2$O$_{8+\delta}$~\cite{Li2007PRL}.

\subsection{Magnetization Properties and Critical Fields}

\begin{figure}[htbp]
   \centering
    \includegraphics[width=\linewidth]{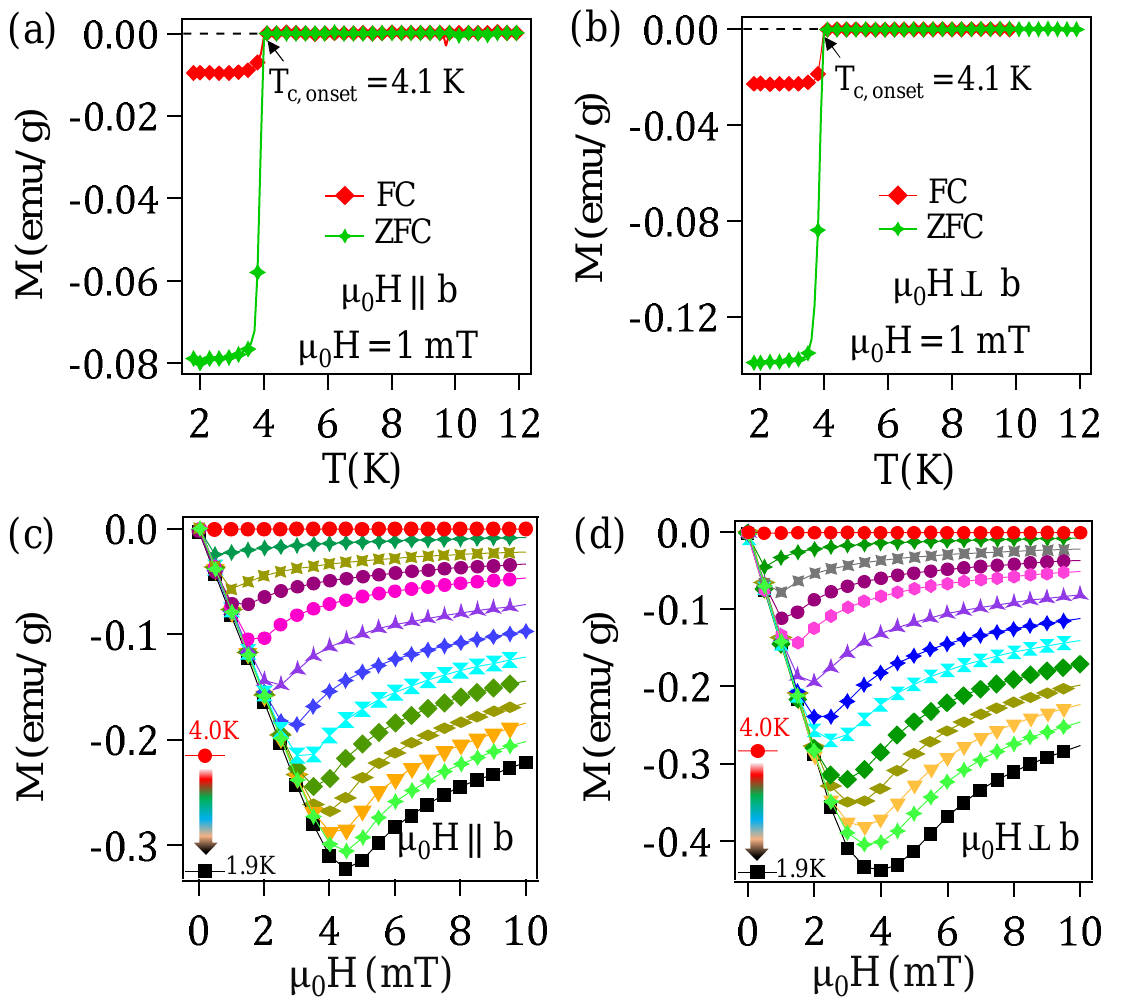}
\caption{Temperature dependence of the magnetization measured in both zero-field-cooled (ZFC) and field-cooled (FC) modes under an applied magnetic field of $\mu_0H=1$~mT for (a) $H\parallel b$ and (b) $H\perp b$ configurations for NiBi$_{2.8}$Pb$_{0.2}$. Temperature-dependent isothermal magnetization curves measured at different temperatures for (c) $H\parallel b$ and (d) $H\perp b$ configurations.}
    \label{fig.4}
\end{figure}

Figs.~\ref{fig.4}(a) and \ref{fig.4}(b) show the temperature-dependent magnetization [$M(T)$] of the $\mathrm{NiBi}_{2.8}\mathrm{Pb}_{0.2}$ single crystal measured under an applied magnetic field of $\mu_0H=1$~mT. Measurements were performed with the magnetic field applied parallel ($H\parallel b$) and perpendicular ($H\perp b$) to the crystallographic $b$ axis under both zero-field-cooled (ZFC) and field-cooled (FC) conditions. For both field orientations, a sharp diamagnetic transition is observed at $T_c\approx4.1$~K, consistent with the bulk superconducting transition determined from the resistivity measurements [see Fig.~\ref{fig.2}(a)]. In contrast, no discernible diamagnetic anomaly is detected near the higher resistive onset temperature, $T_{c,\mathrm{onset}}^{*}\approx8.1$~K. Such behavior is consistent with superconductivity confined to a thin surface or interfacial region, whose small superconducting volume fraction may produce a resistive signature while yielding a magnetic screening signal too weak to be resolved in bulk magnetization measurements~\cite{Li2007PRL,Guo_2020,d1996evidence}. The simultaneous observation of zero resistance and a clear Meissner response below $T_c\approx4.1$~K provides direct evidence for bulk superconductivity~\cite{Jha_2014}. The absence of a corresponding magnetic anomaly near $8.1$~K, despite the distinct resistive onset at this temperature, therefore indicates that the higher-temperature superconducting contribution is non-bulk in nature and supports its assignment to a surface-confined superconducting state~\cite{Li_2007,Li2007PRL}.

\begin{figure}[htbp]
\centering
\includegraphics[width=\linewidth]{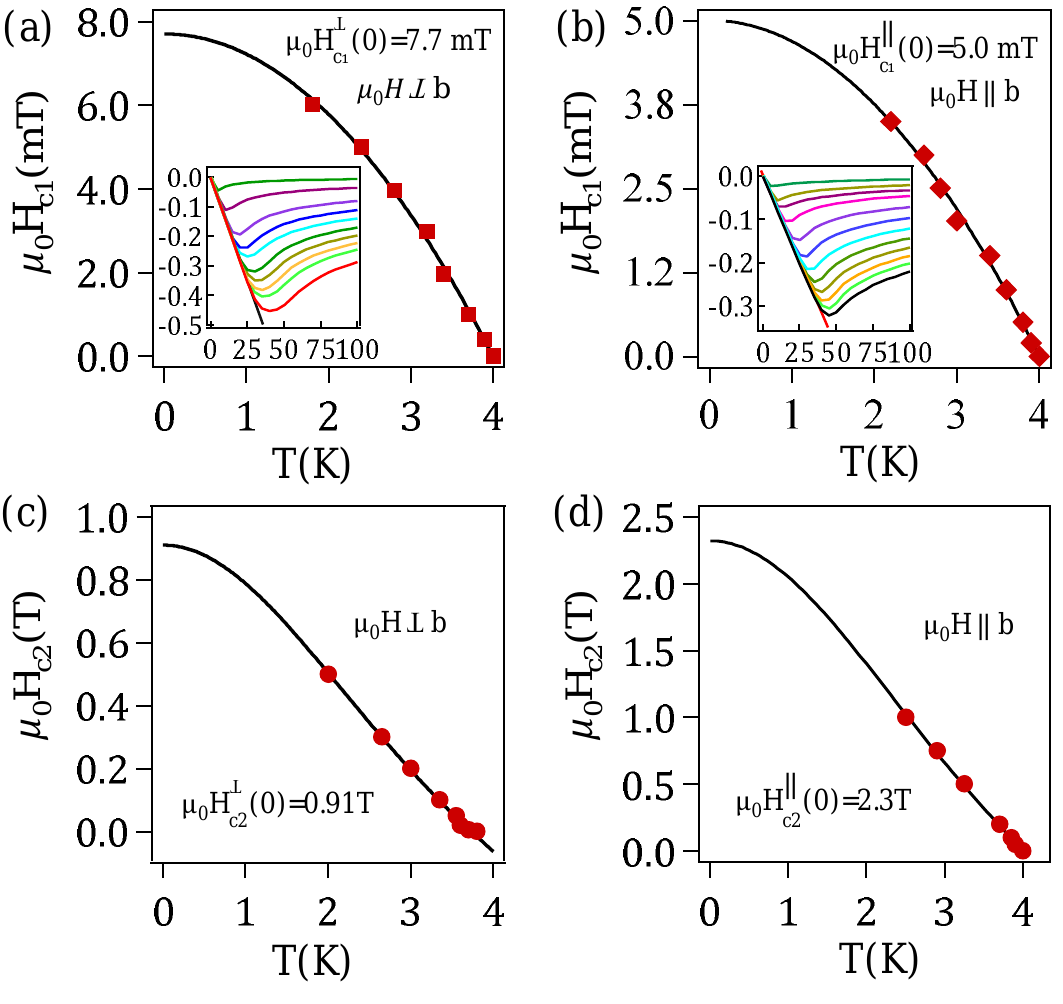}
\caption{Temperature-dependent lower critical field, $\mu_0H_{c1}(T)$, of $\mathrm{NiBi}_{2.8}\mathrm{Pb}_{0.2}$ for
(a) $H\perp b$ and (b) $H\parallel b$. The values were determined from the deviation of the low-field magnetization curves from the initial linear Meissner response. The insets show the corresponding low-field isothermal magnetization curves measured at different temperatures. The solid curves are fits using the Ginzburg--Landau (GL) expression described in the text. Temperature-dependent upper critical field, $\mu_0H_{c2}(T)$, for (c) $H\perp b$ and (d) $H\parallel b$, respectively. The values were determined from the resistive transitions using the 50\% normal-state resistivity criterion. The solid curves are fits using the empirical Ginzburg--Landau expression.}
\label{fig.5}
\end{figure}

\begin{figure}[htbp]
\centering
\includegraphics[width=0.5\linewidth]{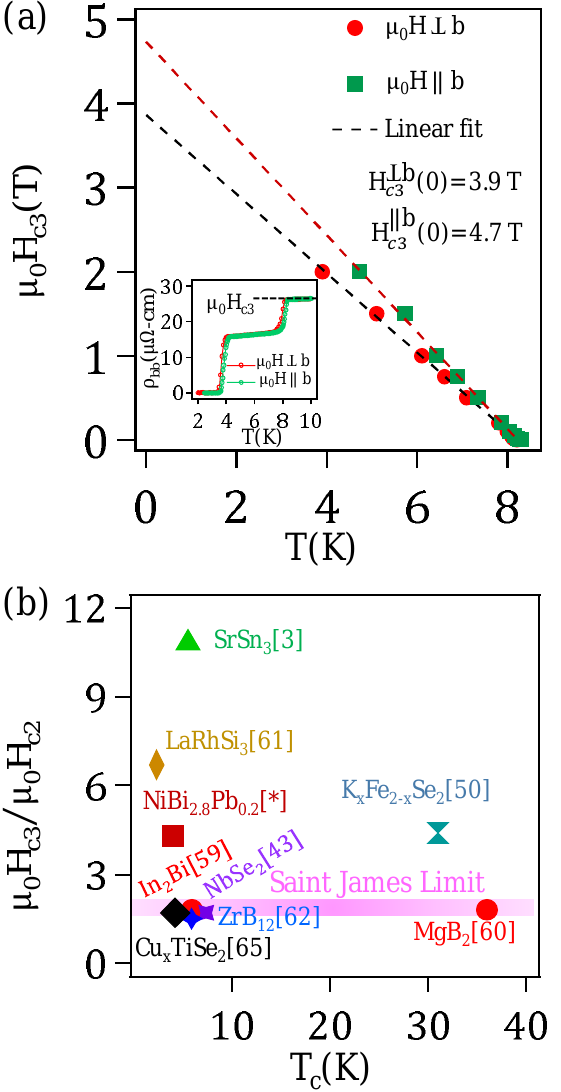}
\caption{(a) Temperature dependence of the surface critical field [$\mu_0H_{c3}(T)$], extracted from the onset of the resistive transitions for both field orientations. Below $T_{c,\mathrm{onset}}^{*}\approx8.1$~K, the data are well described by linear fits, as discussed in the text. (b) Comparison of the surface-to-bulk critical-field ratio [$\mu_0H_{c3}/\mu_0H_{c2}$] and superconducting transition temperature [$T_c$] for selected superconductors. The pink horizontal band denotes the conventional Saint-James--de Gennes limit of $\mu_0H_{c3}/\mu_0H_{c2}=1.695$.}
\label{fig.6}
\end{figure}

The superconducting shielding fraction of the bulk transition is estimated to be nearly $100\%$ from the $4\pi\chi_{\mathrm{eff}}$ $vs.$ $T$ plot [see Fig.~S1 in the Supplemental Material (SM)]. Here,  $\chi_{\mathrm{exp}}=M_{\mathrm{exp}}/H$ is the experimentally measured susceptibility. The demagnetization-corrected susceptibility is given by

\begin{equation}
\chi_{\mathrm{eff}}= \frac{\chi_{\mathrm{exp}}}{1-N\chi_{\mathrm{exp}}},
\end{equation}

where $N$ is the demagnetization factor~\cite{Yan2019PRB,Sundar2019,Lue2013,Stoner_1945}. See Supplementary Materials for details on the determination of $N$. In contrast to the ZFC data, the FC magnetization exhibits a much weaker diamagnetic response, which is attributed to strong magnetic-flux pinning that suppresses flux expulsion during field cooling~\cite{Brandt_1999}.

Figs.~\ref{fig.4}(c) and \ref{fig.4}(d) present the isothermal magnetization [$M(H)$], measured between 1.8 and 4.0~K for $H\parallel b$ and
$H\perp b$, respectively. As expected for a type-II superconductor, the low-field magnetization exhibits an approximately linear field dependence, corresponding to the Meissner state in which magnetic flux is predominantly expelled from the sample~\cite{Song2025PRB,tsindlekht2011superconducting}. With increasing field, $M(H)$ gradually deviates from this initial linear
behavior, marking the onset of vortex penetration at the lower critical field, $\mu_0H_{c1}$. Above $H_{c1}$, magnetic vortices progressively penetrate the sample, leading to the mixed state and characteristic hysteretic $M(H)$ loops [see Fig.~S1(c) in SM].
The finite width of the hysteresis loops indicates irreversible vortex dynamics and substantial vortex pinning~\cite{Brandt_1999}.  The gradual reduction of the hysteresis-loop area with increasing temperature reflects the weakening of vortex pinning as the superconducting state approaches $T_c$.

Figs.~\ref{fig.5}(a) and \ref{fig.5}(b) show the temperature dependence of the lower critical field, $\mu_0H_{c1}(T)$, for $H\perp b$ and
$H\parallel b$, respectively. At each temperature, the field at which the virgin $M(H)$ curve deviates from its initial linear Meissner response was defined as the first penetration field, $\mu_0H_{c1}^{*}$ [see inset of Figs.~\ref{fig.5}(a) and \ref{fig.5}(b) ]. The intrinsic lower critical field was obtained by applying the demagnetization correction,

\begin{equation}
\mu_0H_{c1}(T)= \frac{\mu_0H_{c1}^{*}(T)}{1-N},
\end{equation}

where $N$ is the demagnetization factor determined from the sample geometry and the low-field magnetization data~\cite{Chakrabortty2025,Yadav_2009}.  Further details on the demagnetization factor calculations are provided in the Supplementary Material.

To determine the zero-temperature lower critical field,$\mu_0H_{c1}(0)$, the $\mu_0H_{c1}(T)$ data were fitted using the Ginzburg--Landau (GL) expression~\cite{Shang2023fully,Song2025PRB},

\begin{equation}
\mu_0H_{c1}(T)= \mu_0H_{c1}(0) \left[1-\left(\frac{T}{T_c}\right)^2\right].
\end{equation}

The best GL fits, shown in Figs.~\ref{fig.5}(a) and \ref{fig.5}(b), yield $\mu_0H_{c1}^{\perp b}(0)=7.7$~mT and $\mu_0H_{c1}^{\parallel b}(0)=5.0$~mT, respectively. These values are comparable to the lower critical fields previously reported for the parent compound $\mathrm{NiBi}_3$~\cite{Fujimori2000,Shang2023fully}.

Figs.~\ref{fig.5}(c) and \ref{fig.5}(d) show the temperature dependence of the upper critical field, $\mu_0H_{c2}(T)$, for $H\perp b$ and
$H\parallel b$, respectively. The values of $\mu_0H_{c2}(T)$ were determined from the resistivity measurements shown in Figs.~\ref{fig.2}(c) and \ref{fig.2}(d), using the criterion of $50\%$ of the normal-state resistivity~\cite{Chauhan_2022}. To estimate the zero-temperature upper critical field [$\mu_0H_{c2}(0)$], the experimental data were fitted using the empirical Ginzburg--Landau (GL) expression~\cite{Purwar_2024,Shang2023fully,Chauhan_2022,Awana_2010},

\begin{equation}
\mu_0H_{c2}(T)=\mu_0H_{c2}(0)\left(\frac{1-t^2}{1+t^2}\right),
\end{equation}

where $t=T/T_c$ is the reduced temperature. The best fits, shown in Figs.~\ref{fig.5}(c) and \ref{fig.5}(d), yield $\mu_0H_{c2}^{\perp b}(0)=0.91$~T and $\mu_0H_{c2}^{\parallel b}(0)=2.3$~T, respectively. For comparison, the corresponding upper critical fields reported for the parent compound $\mathrm{NiBi}_3$ are $\mu_0H_{c2}^{\perp b}(0)=0.41$~T and $\mu_0H_{c2}^{\parallel b}(0)=0.70$~T~\cite{Shang2023fully}.
Thus, Pb substitution substantially enhances the upper critical field for both field orientations, demonstrating the increased robustness of superconductivity against an applied magnetic field in $\mathrm{NiBi}_{2.8}\mathrm{Pb}_{0.2}$. The upper critical field also exhibits pronounced anisotropy, quantified by

\begin{equation}
\Gamma_{H_{c2}}= \frac{H_{c2}^{\parallel b}(0)}{H_{c2}^{\perp b}(0)} = \frac{2.3}{0.91} \approx2.5.
\end{equation}

This anisotropy reveals a strong directional dependence of the superconducting properties, which may arise from anisotropic coherence lengths, an altered electronic structure upon Pb substitution, and direction-dependent scattering processes. The coexistence of surface and bulk superconducting states may also contribute to the enhanced high-field response of the doped crystal. Despite the enhancement of $H_{c2}$ upon Pb substitution, both  $\mu_0H_{c2}^{\perp b}(0)$ and $\mu_0H_{c2}^{\parallel b}(0)$ remain well below the weak-coupling Pauli
paramagnetic limit, given by

\begin{equation}
\mu_0H_{\mathrm{P}}=1.86T_c.
\end{equation}

For $T_c\approx4.1$~K, the Pauli limiting field is $\mu_0H_{\mathrm{P}}\approx7.6$~T, which is substantially larger than
$\mu_0H_{c2}^{\perp b}(0)=0.91$~T and $\mu_0H_{c2}^{\parallel b}(0)=2.3$~T. These results suggest that the upper critical field of NiBi$_{2.8}$Pb$_{0.2}$ is predominantly governed by orbital pair breaking mechanism~\cite{Shang2023fully,Song2025PRB,Clogston_1962}. Notably, the substantially larger value of $H_{c2}^{\parallel b}$ indicates that superconductivity is more robust against fields applied parallel to the
$b$ axis than against fields applied perpendicular to it.

Within the Ginzburg--Landau (GL) framework, the magnetic penetration depth $\lambda$ is related to the coherence length $\xi$ and the lower critical field $\mu_0H_{c1}$ through~\cite{Kalenyuk_2017,Vedeneev_2013,Tinkham1996,Shang2023fully,Clem_1989}

\begin{equation}
\mu_0H_{c1}=
\frac{\Phi_0}{4\pi\lambda^2}
\ln\left(\frac{\lambda}{\xi}\right)
=
\frac{\Phi_0}{4\pi\lambda^2}\ln\kappa,
\end{equation}

where $\Phi_0=h/(2e)=2.07\times10^{-15}$~T\,m$^2$ is the magnetic flux quantum and $\kappa=\lambda/\xi$ is the GL parameter. For the quantitative determination of $\kappa$, the finite vortex-core correction was included, such that the logarithmic term is replaced by $\ln\kappa+0.5$. The zero-temperature coherence lengths were obtained from the anisotropic GL relations~\cite{Song2025PRB,Chauhan_2022,Vedeneev_2013},

\begin{align}
\xi^{\parallel b}(0) &= \sqrt{\frac{\Phi_0}{2\pi\mu_0H_{c2}^{\perp b}(0)}}, \\
\xi^{\perp b}(0)&=\frac{\Phi_0}{2\pi\xi^{\parallel b}(0)\mu_0H_{c2}^{\parallel b}(0)}.
\label{eq:coherence_lengths}
\end{align}

The resulting coherence lengths are $\xi^{\parallel b}(0)\approx19$~nm and $\xi^{\perp b}(0)\approx7.54$~nm. The corresponding anisotropy parameter is

\begin{equation}
\Gamma=
\frac{\xi^{\parallel b}(0)}{\xi^{\perp b}(0)} \approx2.52.
\end{equation}

This value is substantially larger than the anisotropy of the parent $\mathrm{NiBi}_3$ single crystal. Shang \textit{et al.} reported
$\mu_0H_{c2}^{\parallel b}(0)=0.70$~T and $\mu_0H_{c2}^{\perp b}(0)=0.41$~T, corresponding to an upper-critical-field
anisotropy of $\Gamma_{H_{c2}}\approx1.7$.\cite{Shang2023fully} In contrast, the present Pb-substituted crystal exhibits
$\mu_0H_{c2}^{\parallel b}(0)=2.3$~T and $\mu_0H_{c2}^{\perp b}(0)=0.91$~T, yielding $\Gamma_{H_{c2}}\approx2.5$. Pb substitution therefore enhances the upper-critical-field anisotropy by approximately $50\%$. Using the same GL formalism, the coherence lengths of the parent compound are approximately $\xi^{\parallel b}(0)\approx21.7$~nm and $\xi^{\perp b}(0)\approx28.3$~nm. The marked reduction of both coherence lengths in $\mathrm{NiBi}_{2.8}\mathrm{Pb}_{0.2}$ accounts for the enhanced $\mu_0H_{c2}$ values. In particular, the smaller
$\xi^{\perp b}(0)$ is responsible for the substantially larger $H_{c2}^{\parallel b}$, indicating improved robustness of the superconducting
state against orbital pair breaking for fields applied along the $b$ axis. The GL parameter was estimated from

\begin{equation}
\frac{H_{c1}(0)}{H_{c2}(0)}
\simeq
\frac{\ln\kappa(0)+0.5}{2\kappa^2(0)}.
\label{eq:kappa}
\end{equation}

The resulting values are $\kappa^{\parallel b}(0)\approx30$ and $\kappa^{\perp b}(0)\approx13.5$. Since both values are much larger than
$1/\sqrt{2}$, $\mathrm{NiBi}_{2.8}\mathrm{Pb}_{0.2}$ is clearly a strongly type-II superconductor. The large $\kappa$ values also imply a broad mixed-state region between $H_{c1}$ and $H_{c2}$, consistent with the observed vortex penetration and hysteretic magnetization behavior. The GL penetration depths were determined from

\begin{equation}
\lambda^{\parallel b}(0)=
\kappa^{\perp b}(0)\xi^{\parallel b}(0),
\qquad
\lambda^{\perp b}(0)=
\lambda^{\parallel b}(0)
\frac{\xi^{\parallel b}(0)}{\xi^{\perp b}(0)}.
\label{eq:lambda}
\end{equation}

The resulting penetration depths are $\lambda_{\mathrm{GL}}^{\parallel b}(0)=256$~nm and $\lambda_{\mathrm{GL}}^{\perp b}(0)=646$~nm. The parent $\mathrm{NiBi}_3$ single crystal exhibits much more isotropic penetration depths, with GL estimates of approximately 229~nm and 238~nm for $H\parallel b$ and $H\perp b$, respectively~\cite{Shang2023fully}. Furthermore, transverse-field $\mu$SR measurements on the parent compound yielded penetration depths of 223~nm and 210~nm for the two orientations. Thus, $\lambda_{\mathrm{GL}}^{\parallel b}$ remains comparable to the parent value, whereas $\lambda_{\mathrm{GL}}^{\perp b}$ is enhanced by a factor of
approximately 2.7.

Since $\lambda^{-2}$ is proportional to the superfluid density within the London description, the substantially enhanced GL-derived penetration depth, $\lambda_{\mathrm{GL}}^{\perp b}$, suggests an orientation-dependent reduction of the effective superfluid stiffness upon Pb substitution. At the same time, the lower critical fields do not evolve uniformly.  The pristine NiBi$_3$ has $\mu_0H_{c1}^{\parallel b}(0)\approx7.4$~mT and $\mu_0H_{c1}^{\perp b}(0)\approx6.2$~mT~\cite{Shang2023fully},  whereas the Pb-substituted NiBi$_{2.8}$Pb$_{0.2}$ shows $\mu_0H_{c1}^{\parallel b}(0)=5.0$~mT and
$\mu_0H_{c1}^{\perp b}(0)=7.7$~mT (see Tab.~\ref{tab:SC_parameters} for the summary of extracted superconducting parameters from the parent and Pb-substituted $\mathrm{NiBi}_3$ crystals). Such contrasting trends in $H_{c1}$ and $H_{c2}$ are not unexpected, since $H_{c1}$ is governed predominantly by the penetration depth $\lambda$ (with a weaker logarithmic dependence on $\kappa$), whereas $H_{c2}$ is primarily controlled by the coherence length $\xi$. Direct measurements of the penetration depth by transverse-field $\mu$SR, microwave spectroscopy, or tunnel-diode resonator measurements would be valuable for independently confirming the large penetration-depth anisotropy. The enhanced anisotropy may originate from Pb-induced changes in the electronic structure, anisotropic effective masses, and/or direction-dependent scattering rates.


\begin{table}[t]
\centering
\caption{Superconducting parameters of NiBi$_{2.8}$Pb$_{0.2}$
($T_{\mathrm{c}} \sim 4$ K).}
\label{tab:SC_parameters}

\renewcommand{\arraystretch}{1.5}

\begin{tabular}{|l|c|c|c|c|}
\hline
&
\multicolumn{2}{c|}{$\mu_0 H \parallel b$}
&
\multicolumn{2}{c|}{$\mu_0 H \perp b$} \\

\cline{2-5}

\textbf{Parameter}
& \text{NiBi$_3$~\cite{Shang2023fully}}
& \text{NiBi$_{2.8}$Pb$_{0.2}$}
& \text{NiBi$_3$~\cite{Shang2023fully}}
& \text{NiBi$_{2.8}$Pb$_{0.2}$} \\

\hline

$\mu_0 H_{\mathrm{c1}}(0)$ (mT)
& 7.4 & 5 & 6.2 & 7.7 \\

$\mu_0 H_{\mathrm{c2}}(0)$ (T)
& 0.7 & 2.3 & 0.41 & 0.91 \\

$\xi(0)$ (nm)
& 21.7 & 19 & 28.3 & 7.54 \\

$\kappa(0)$ & 10.5 & 30 & 8.4 & 13.5 \\

$\lambda_{\mathrm{GL}}(0)$ (nm)
& 229 & 256.5 & 238 & 646 \\

\hline
\end{tabular}
\end{table}

The robustness of the surface superconducting state under an applied magnetic field can be quantified by the ratio of the surface critical field, $H_{c3}$, to the bulk upper critical field, $H_{c2}$~\cite{kuang2021magnetization,Song2025PRB}. The surface critical field was determined from the onset of the resistive transition at each applied field [see Figs.~\ref{fig.2}(c) and \ref{fig.2}(d)]. Fig.~\ref{fig.6}(a) shows the temperature dependence of $H_{c3}$ for both field orientations. In both cases, $H_{c3}(T)$ increases
linearly upon cooling over the measured temperature range, with no clear indication of saturation. Linear extrapolation to $T=0$ yields
$\mu_0H_{c3}^{\parallel b}(0)=4.7$~T and $\mu_0H_{c3}^{\perp b}(0)=3.9$~T. Using the bulk upper critical fields determined from the $50\%$ resistivity criterion, the corresponding ratios are

\begin{equation}
\frac{H_{c3}^{\parallel b}}{H_{c2}^{\parallel b}} = \frac{4.7}{2.3} \approx2.04,
\qquad
\frac{H_{c3}^{\perp b}}{H_{c2}^{\perp b}}= \frac{3.9}{0.91} \approx4.30.
\end{equation}

Both values exceed the Saint-James--de Gennes limit, $H_{c3}/H_{c2}\approx1.695$, expected for a conventional single-band superconductor with an ideal surface and a magnetic field parallel to that surface~\cite{lyard2002anisotropy,kimura2016type,tsindlekht2004tunneling}.
Within conventional GL theory, surface superconductivity survives up to
\begin{equation}
H_{c3}(T)\approx1.695H_{c2}(T).
\end{equation}

The ratio of approximately 2.04 for $H\parallel b$ therefore already indicates an enhanced surface superconducting response relative to the conventional GL expectation. The much larger ratio of approximately 4.30 for $H\perp b$ suggests an exceptionally robust high-field surface contribution~\cite{Eltsev2002}. However, the Saint-James--de Gennes limit applies most directly to a semi-infinite sample with a field parallel to an ideal planar surface. Therefore, the perpendicular-field ratio should be interpreted with care because finite-size effects, sample geometry, surface roughness, field misalignment, and orientation-dependent resistive broadening can modify the experimentally extracted value of $H_{c3}/H_{c2}$.

It is also important to note that $H_{c3}$ and $H_{c2}$ are extracted using different resistive criteria:  $H_{c3}$ is determined from the transition onset, whereas $H_{c2}$ is obtained using the $50\%$ normal-state-resistivity criterion. The reported ratios should therefore be regarded as qualitative measured that characterize the high-field transport response. Nevertheless, their systematic enhancement, together with the cusp-like angular dependence of $H_{c3}$, provides strong evidence for a robust surface-confined superconducting state.
A possible origin of $H_{c3}/H_{c2}>1.695$ is a modification of the boundary condition for the superconducting order parameter. This effect can be described by the de Gennes extrapolation length $b$,

\begin{equation}
b=\left.\frac{\psi(x)}{\partial\psi(x)/\partial x}\right|_{x=0},
\end{equation}

where $\psi(x)$ is the superconducting order parameter and $x$ is the coordinate normal to the surface. Under this sign convention, a finite negative extrapolation length ($b<0$) corresponds to an enhancement of the order parameter near the surface and can lead to a surface critical temperature higher than the bulk transition temperature~\cite{fink1969surface, kimura2016type, Song2025PRB}. The unusually large $H_{c3}$ values observed in the present crystal may therefore reflect modified surface boundary conditions, possibly associated with Pb-induced changes in the near-surface electronic structure, local stoichiometry, disorder, or strain.

These results should be distinguished from the enhanced bulk $H_{c2}$ values.  $H_{c2}$ characterizes the destruction of bulk superconductivity, whereas $H_{c3}$ describes the persistence of superconductivity in a surface or interfacial region for $H_{c2}<H<H_{c3}$. Thus, Pb substitution appears to enhance both the anisotropic bulk field robustness and the stability of the surface superconducting state, with the latter producing a distinct high-field, quasi-two-dimensional superconducting regime.

\begin{figure}[htbp]
\centering
\includegraphics[width=\linewidth]{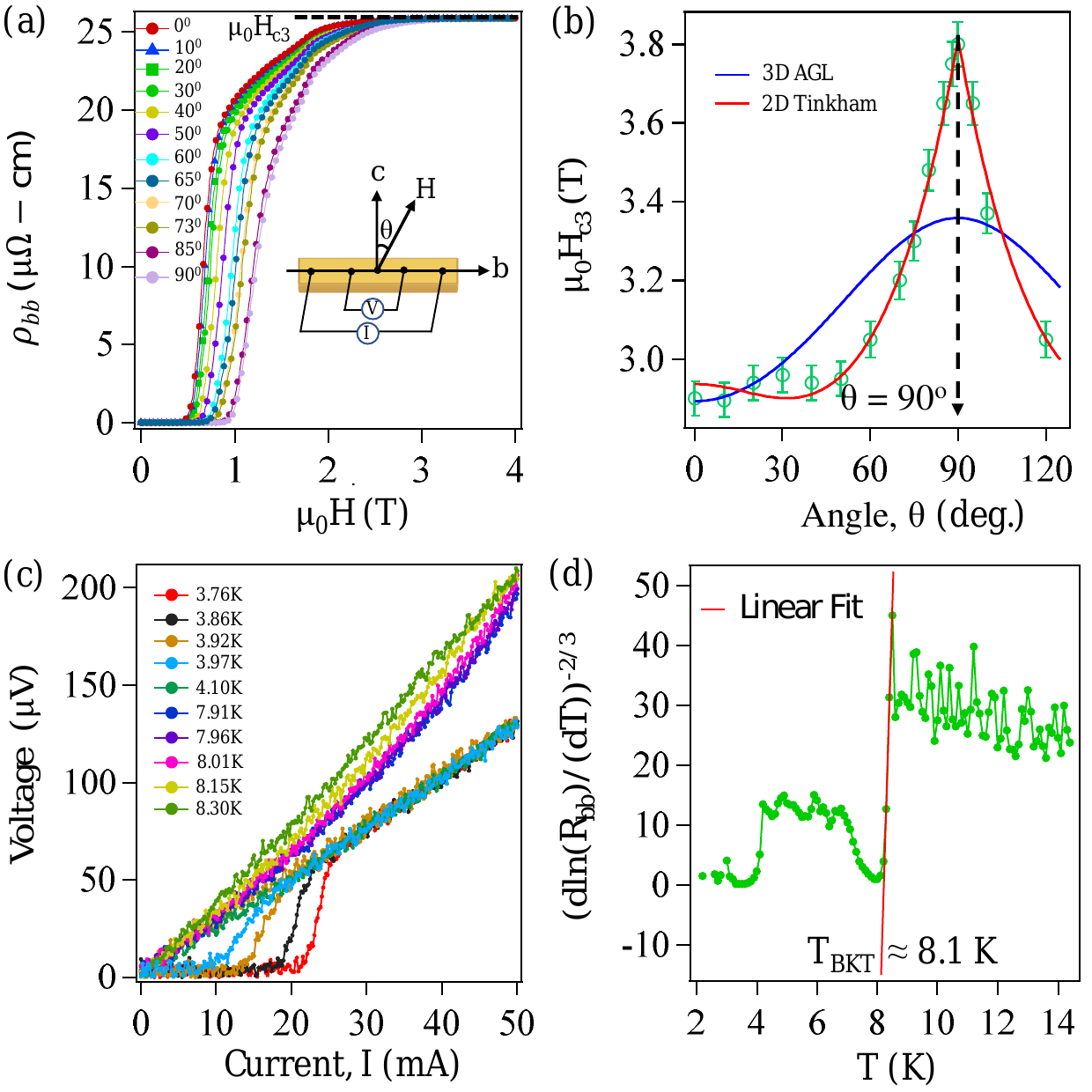}
\caption{(a) Field-dependent in-plane resistivity, $\rho_{bb}(\mu_0H)$, measured at 2~K for different field angles ($\theta$). The inset illustrates the measurement geometry and the definition of $\theta$. (b) Angular dependence of the surface critical field [$\mu_0H_{c3}(\theta)$] is analyzed using the anisotropic three-dimensional Ginzburg--Landau (3D AGL) model and the two-dimensional (2D) Tinkham model. (c) Current--voltage ($I$--$V$) characteristics measured at different temperatures across the superconducting transition. (d)
Halperin--Nelson analysis, $(d\ln R_{bb}/dT)^{-2/3}$ versus $T$, used to determine the Berezinskii--Kosterlitz--Thouless transition temperature, $T_{BKT}\approx8.1$~K. The linear behavior provides evidence for two-dimensional surface superconductivity.}
\label{fig.7}
\end{figure}

\subsection{Two-Dimensional Nature of Surface Superconductivity}

To investigate the nature of the superconducting transition at $T_{c,\mathrm{onset}}^{*}=8.1$~K, magnetic-field-dependent resistivity
measurements were performed at 2~K for different field orientations ($\theta$) [see Fig.~\ref{fig.7}(a)]. Here, $\theta$ denotes the angle between the sample-surface normal and the applied magnetic field. As illustrated in the inset of Fig.~\ref{fig.7}(a), the surface critical field, $\mu_0H_{c3}$, was determined from the onset of the $R(\mu_0H)$ transition at each field orientation. The resulting angular dependence of $\mu_0H_{c3}$ at 2~K is shown in Fig.~\ref{fig.7}(b). The surface-superconducting transition shifts systematically to higher magnetic fields as the field is rotated from the out-of-plane configuration ($\theta=0^\circ$) to the in-plane configuration ($\theta=90^\circ$). A pronounced cusp-like peak near $\theta\approx90^\circ$ indicates a strongly anisotropic superconducting response characteristic of a two-dimensional (2D) superconducting state. In the field range $\mu_0H_{c2}<\mu_0H<\mu_0H_{c3}$, bulk superconductivity is suppressed, whereas superconductivity can persist within a thin surface or interfacial region. This confinement gives rise to the sharp angular dependence of the critical field and the cusp near $\theta\approx90^\circ$.\cite{Zeinali_2016,Shen_2020,levy2016puzzling}

To quantify the angular dependence, the experimental $\mu_0H_{c3}(\theta)$ data were fitted using both the anisotropic three-dimensional
Ginzburg--Landau (3D AGL) model and the two-dimensional Tinkham model~\cite{Morris1972,Tinkham1996,Mineev2001, Kittaka2009}. The 3D AGL model assumes coherent superconductivity along all three crystallographic directions, with finite interlayer coupling and direction-dependent coherence lengths. It therefore predicts a smooth, elliptical angular dependence~\cite{Blatter1992},

\begin{equation}
\left(\frac{H_{c3}(\theta)\cos\theta}{H_{c3}^{\perp}}\right)^2+\left(\frac{H_{c3}(\theta)\sin\theta}{H_{c3}^{\parallel}}\right)^2=1,
\end{equation}

where $H_{c3}^{\perp}=H_{c3}(0^\circ)$ and $H_{c3}^{\parallel}=H_{c3}(90^\circ)$. In contrast, the 2D Tinkham model applies when superconductivity is confined to a thin or quasi-2D layer with weak interlayer coupling. The angular dependence is then described by

\begin{equation}
\left|
\frac{H_{c3}(\theta)\cos\theta}{H_{c3}^{\perp}}
\right|
+
\left(
\frac{H_{c3}(\theta)\sin\theta}{H_{c3}^{\parallel}}
\right)^2
=1.
\end{equation}

The absolute-value term produces a cusp at $\theta=90^\circ$, in contrast to the smooth angular variation predicted by the 3D AGL model~\cite{fink1969surface, Saito_2016, Patra2022}. The excellent agreement between the experimental data and the 2D Tinkham fit [Fig.~\ref{fig.7}(b)] supports the two-dimensional character of the surface-superconducting transition at $T_{c,\mathrm{onset}}^{*}\approx8.1$ K. The fit yields $\mu_0H_{c3}^{\perp}=2.9$~T and $\mu_0H_{c3}^{\parallel}=3.8$~T, in close agreement with the corresponding transition fields extracted independently from the $R(T)$ measurements under different magnetic fields [Fig.~\ref{fig.6}(a)]. Fig.~\ref{fig.7}(c) shows the in-plane current--voltage ($I$--$V$) characteristics measured at different temperatures. At low temperatures, a distinct zero-voltage plateau is observed up to the critical current ($I_c$),  confirming the presence of superconductivity. The critical current decreases systematically with increasing temperature and vanishes near the bulk superconducting transition temperature, $T_c\approx4.1$~K. Above 4.1~K, where only the surface superconducting state remains, the $I$--$V$ curves become strongly nonlinear, consistent with the expected transport behavior of surface superconductivity~\cite{Zeinali_2016}. Upon further heating, a linear Ohmic
response is recovered, indicating the transition to the normal state.

The two-dimensional nature of the superconducting state was further examined through the Berezinskii--Kosterlitz--Thouless (BKT) transition~\cite{Shen_2020, Guo_2020}. For a 2D superconductor, the resistivity just above the BKT transition temperature follows the Halperin--Nelson relation~\cite{ Halperin1979},

\begin{equation}
\rho(T)=\rho_n\exp\left[-b\,t^{-1/2}\right],
\end{equation}

where $\rho_n$ and $b$ are fitting parameters, and $t=T/T_{\mathrm{BKT}}-1$ is the reduced temperature~\cite{Saito_2016}.
As shown in Fig.~\ref{fig.7}(d), the high-temperature superconducting transition is well described by the Halperin--Nelson model, yielding
$T_{\mathrm{BKT}}=8.1$~K close to  $T_{c,\mathrm{onset}}^{*}\approx8.1$ K. This agreement provides strong evidence that the emergent superconducting phase is a two-dimensional surface superconducting state.

\section{Conclusion}

In summary, we demonstrate that Pb substitution in the conventional superconductor NiBi$_3$ induces a high-temperature surface superconducting state with an onset temperature of $T_{c,\mathrm{onset}}^{*}\approx8.1$ K, while preserving the intrinsic bulk superconductivity at $T_c\approx4.1$ K. Structural characterization confirms Pb incorporation without altering the crystal symmetry, while complementary resistivity, magnetization, specific-heat, and angle-dependent magnetotransport measurements reveal distinct surface and
bulk superconducting states. The absence of a discernible Meissner response and specific-heat anomaly near  $T_{c,\mathrm{onset}}^{*}\approx8.1$ K supports a non-bulk origin of the higher-temperature transition, whereas the cusp-like
angular dependence of $H_{c3}$, nonlinear $I$--$V$ characteristics, and BKT behavior provide strong evidence for the two-dimensional nature of the surface superconducting state. Remarkably, the surface-to-bulk critical-field ratio reaches $H_{c3}/H_{c2}\approx4.3$ for $H\perp b$, substantially exceeding the conventional Saint-James--de Gennes value of 1.695. These findings establish Pb substitution as an effective route for engineering robust surface superconductivity in NiBi$_3$ and provide a platform for exploring the coexistence of bulk and low-dimensional superconducting states.

\section{Experimental Section}

Single crystals of NiBi$_3$ and Pb-substituted NiBi$_3$ were grown using a self-flux method with excess Bi as the flux~\cite{Zhu2012surface,Shang2023fully}. In this method, high-purity Ni powder (99.9\%, Alfa Aesar), Bi pieces (99.99\%, Alfa Aesar), and Pb pieces (99.9\%, Alfa Aesar) were mixed in a nominal molar ratio of Ni:Bi:Pb $=1:(10-x):x$. The starting materials were loaded into an alumina crucible, sealed in an evacuated quartz ampoule, and heated to $1050^{\circ}\mathrm{C}$ for 24~h to ensure complete homogenization of the melt. The furnace was subsequently cooled slowly to $360^{\circ}\mathrm{C}$ at a rate of $2^{\circ}\mathrm{C/h}$. At $360^{\circ}\mathrm{C}$, the excess Bi flux was removed by centrifugation. This procedure yielded needle-like NiBi$_3$ single crystals with typical dimensions of approximately $3\times0.3\times0.3~\mathrm{mm^3}$, whereas the Pb-substituted crystals were comparatively smaller.

The phase purity and crystallographic orientation of the crystals were characterized by X-ray diffraction (XRD) technique using a Rigaku SmartLab diffractometer equipped with a 9-kW Cu $K_\alpha$ X-ray source. Powder XRD measurements were performed on crystals crushed from the as-grown samples. The chemical compositions of the as-grown crystals were determined by energy-dispersive X-ray spectroscopy (EDS). Electrical resistivity and magnetotransport measurements were performed using a standard four-probe configuration. Electrical transport,  magnetotransport, and specific heat measurements were carried out in a 9-T Physical Property Measurement System (PPMS, DynaCool, Quantum Design). Magnetic measurements were carried out using the 7-T Magnetic Property Measurement System (MPMS). High-resolution transmission electron microscopy (HRTEM) was performed using an FEI Tecnai TF20 S-TWIN microscope operated at an accelerating voltage of 200~kV.

\section*{Acknowledgements}
S.T. thanks the Science and Engineering Research Board (SERB) and the Anusandhan National Research Foundation (ANRF), India, for financial support through Grant Nos. CRG/2023/00748 and ANRF/ARG/2025/007085/PS, respectively. This research utilized the Technical Research Centre (TRC) instrumentation facilities at the S. N. Bose National Centre for Basic Sciences, established under the TRC project of the Department of Science and Technology (DST), Government of India. We acknowledge the UGC-DAE Consortium for Scientific Research, Indore, for providing access to the 7-T Magnetic Property Measurement System (MPMS) facility.

\bibliographystyle{elsarticle-num}
\bibliography{NiBi3Pb}

\end{document}